\documentstyle[12pt]{article}
\begin{document}

\begin{center}
{\LARGE{\bf OFF SHELL  $\pi N$  AMPLITUDE\\
\vspace{0,3cm}
AND THE $pp \rightarrow pp \pi^{0}$ REACTION\\
\vspace{0,3cm}
NEAR THRESHOLD}}
\end{center}

\begin{center}
{\large{E. Hern\'andez and E. Oset.}}
\end{center}

\begin{center}
{\small{\it  Departamento de F\'{\i}sica Te\'orica and IFIC\\
Centro Mixto Universidad de Valencia - CSIC\\
46100 Burjassot (Valencia) Spain.}}
\end{center}

\vspace{3cm}

\begin{abstract}
{\small{We have used a conventional  model for the $pp \rightarrow pp \pi^{0}$
reaction consisting of the Born term plus the $s$-wave rescattering term. As
a novelty we have introduced the off shell dependence of the $\pi N$
$s$-wave isoscalar amplitude. This amplitude is appreciably enhanced when one
moves to the off shell situations met in the problem and, as a consequence,
the $pp \rightarrow pp \pi^{0}$ cross section becomes considerably larger than
with the use of the $\pi N$ on shell amplitudes. Two different models for the
off shell extrapolation found in the literature have been used and the cross
sections obtained are large enough to account for the experimental data,
although
uncertainties remain due to the incomplete knowledge of the off shell
extrapolation.}}
\end{abstract}

\newpage
\baselineskip 0.7cm

The $\pi N$ $s$-wave amplitude is commonly derived from an effective
Hamiltonian \cite{1}

\begin{equation}
H= 4 \pi \frac{\lambda_{1}}{m_{\pi}} \bar{\Psi} \vec{\phi} \vec{\phi} \Psi +
4 \pi \frac{\lambda_{2}}{m_{\pi}^{2}} \bar{\Psi} \vec{\tau}. \vec{\phi} \times
\vec{\Pi} \Psi
\end{equation}

\noindent
and $\lambda_{1}, \lambda_{2}$ are related to the scattering lengths. A
remarkable feature of this interaction is the large cancellation between
amplitudes which leads to very small values of $\lambda_{1}$ compared to
$\lambda_{2}$. By using H\"ohler's results \cite{2} one obtains
$\lambda_{1} = 0.0075, \lambda_{2}= 0.053$ for on shell $\pi N$ scattering at
threshold.

One of the consequences of the smallness of $\lambda_{1}$ is the serious
disagreement with the data \cite{3} of the computed cross section for
$pp \rightarrow pp \pi^{0}$, based on the Born term and the rescattering terms
of fig. 1 using the threshold on shell $\pi N$ scattering amplitude
\cite{4,5}.

A plausible solution was given in \cite{6}, based on a relativistic
description of the $NN$ interaction which generates pair terms where the
$N \bar{N}$ components are connected by the isoscalar part of the interaction.

Although this procedure is univocally defined when dealing with $N$ positive
energy components \cite{7,8}, this is not the case when extrapolating the
results to negative energy components where large
ambiguities are
present. As an example the authors of ref. \cite{9}, who try to get the
relativistic potentials from information of amplitudes involving both
nucleons and antinucleons, get a strength for the vector and scalar potentials
of about half the results in \cite{8}. Furthermore, modern versions of the
$NN$ potential \cite{10} where the intermediate range attraction ($\sigma$
exchange in most models) is obtained explicitly from the exchange of two
correlated pions, would lead to different results in the $\bar{N}N$ sector
which would reduce the effects found in \cite{6} if pseudovector couplings are
used for the $NN\pi$ vertex.

The purpose of the present work is to provide an alternative explanation
for the
unexpectedly large $pp \rightarrow pp \pi^{0}$ cross section. The idea is based
on the findings of ref. \cite{11}, where $s$-wave pion absortion for pionic
atoms was studied. It was found there that, because in the rescattering term
of fig. 1b (for an incoming pion) the $\pi N$ $s$-wave amplitude appears half
off shell, the term proportional to $\lambda_{1}^{2}$ in the absortion rate
was appreciably enhanced due to the off shell amplitude met in the process,
which is considerably larger than the on shell one. One should also
note that because the absortion rate for charged pions is dominated by
a term proportional to $\lambda_2^2$, the effects due to the off shell
extrapolation of $\lambda_1$ were small \cite{11} and  absortion experiments
(of charged pions) never felt the need for it \cite{12}.

The increase of the isoscalar amplitude as one moves off shell is a feature
 of all known models for the off shell extrapolation which
can be derived from basic arguments using constraints of current algebra and
PCAC \cite{13}. This is also the
case in a recent model for the $\pi N$ scattering which treats again the
$\sigma$ exchange as a correlated two pion exchange \cite{14}.

The amplitudes corresponding to figs. 1a, 1b close to pion threshold are
given, in Mandl Shaw normalization \cite{15}, by

\begin{equation}
-i t^{(a)} (p_{1} p_{2}, p'_{1} p'_{2} p_{\pi}) = \frac{f}{m_{\pi}}
\frac{p_{\pi}^{0}}{2M} < s'_{2} |\vec{\sigma}_{2} (\vec{p}_{2} + \vec{p}_{2}
\, ')|s_{2}> + (1 \leftrightarrow 2) + exchange
\end{equation}

$$
-i t^{(b)} (p_{1} p_{2}, p'_{1} p'_{2} p_{\pi}) =
- \frac{f}{m_{\pi}} <s'_{1}
|\vec{\sigma} \vec{q}|s_{1}> F (q) \frac{i}{q^{2}-m_{\pi}^{2}}
$$

\begin{equation}
< s'_{2}|(-i) 4 \pi \frac{2\lambda_{1} (q, p_{\pi})}{m_{\pi}}|s_{2} > +
(1 \longleftrightarrow 2) + exchange
\end{equation}

\noindent
where the exchange is done in the two incoming protons (a symmetry factor
$\frac{1}{2}$ will appear in the standard formula for the integrated cross
section because of the identity of the two final protons). F(q) is the
monopole form factor with $\Lambda = 1250 \, MeV$ from \cite{16}.

The isoscalar $\pi N$ amplitude $\lambda_{1} (q, p_{\pi})$ appears half off
mass shell, $q^{0} = 70 \, MeV, q = 370 MeV/c$. At this off shell momenta this
amplitude is, in most models, about five times bigger than the on shell
amplitude.

We have taken two different off shell extrapolations to get an idea of the
uncertainties that one can have in the final cross section.

The first model is due to Hamilton and the $\pi N$ isoscalar amplitude is due
to $\sigma$-exchange plus a short range piece \cite{17}. In this model we
have

\begin{equation}
\lambda_{1} (q, p_{\pi})= - \frac{1}{2} (1 + \epsilon ) m_{\pi} \left[
a_{sr} + a_{\sigma} \frac{m^{2}_{\sigma}}{m_{\sigma}^{2} - (q-p_{\pi})^{2}}
\right]
\end{equation}

\noindent
with $\epsilon = m_{\pi}/M, a_{\sigma} = 0.220 \, m_{\pi}^{-1}, a_{sr} =
-0.233 \, m_{\pi}^{-1}, m_{\sigma} = 550 \, MeV$.

The second model is the one of ref. \cite{18}, where it is shown that for not
too large values of the kinematical variables, the isoscalar amplitude
satisfying current algebra constraints can be written as

$$
\overline{F}^{+} (\nu , t; q^{2}, p_{\pi}^{2}) =(\frac{t}{m_{\pi}^{2}} -1)
\frac{\sigma (t)}{f_{\pi}^{2}}
$$

\begin{equation}
+ (p_{\pi}^{2} + q^{2} - t) \frac{\overline{F}^{+} (0, m^{2}_{\pi};
m^{2}_{\pi},
m^{2}_{\pi})}{m_{\pi}^{2}} + f_{3}^{+} \nu^{2}
\end{equation}

\noindent
with

$$
t= (q- p_{\pi})^{2}, \nu = (q + p_{\pi}) (p + p')/4M
$$

\noindent
where $p = (M+ m_{\pi}/2, - \vec{q}),p' = (M,\vec{0}),$
are the initial and final
nucleon momenta.
The first term on the right hand side of the equation is the Adler term
and $\sigma (t)$ is the $\pi N$ sigma commutator given by

\begin{equation}
\sigma (t) = \frac{\sigma (0)}{(1- \frac{t}{m_{1}^{2}})^{2} (1-\frac{t}
{m_{2}^{2}})}
\end{equation}

\noindent
with $m_{1} = 8.24 m_{\pi} \; , \; m_{2} = 7.5 m_{\pi} \; , \sigma (0) =
25 MeV, f_{\pi}= 93 MeV, f_{3}^{+} = 0.82 m_{\pi}^{-3}$ and
$\overline{F}^{+} (0, m_{\pi}^{2};
m_{\pi}^{2}, m_{\pi}^{2}) = - 0.30 m_{\pi}^{-1}$
\cite{18,19}. The normalization of the amplitude in eq. (5) is such that for
on shell pions at threshold

\begin{equation}
\overline{F}^{+} (\nu , t; q^{2}, p_{\pi}^{2}) \equiv
- \frac{4\pi 2 \lambda_{1}}{m_{\pi}}(q, p_{\pi})
\end{equation}

As we said, the values of $\overline{F}^{+}$ needed correspond to half off
shell
situations with $q \simeq 370 \; MeV/c$. Eq. (5) for higher momenta would
become progressively inaccurate \cite{18,19}, but these momenta do not play a
role in the cross section. For this reason, and in order to have well behaved
Fourier transforms in our computational scheme, we regularize
$\overline{F}^{+}$
multiplying it by the function $exp \; (- \vec{q}^{\, 2}/11 m_{\pi}^{2})$ which
does not affect the region of momenta of relevance to the problem. The results
are rather insensitive to this regularizing factor. Indeed, if we change
$\overline{F}^{+}$ at momenta $q$ above $q=500 \, MeV/c$,
making it fall down much
faster for instance, the cross sections change only at the level of $2\%$.
The off shell extrapolation of $\overline{F}^{+}$ obtained with
the quoted regularizing
factor ressembles very much the results of model (2) of ref. \cite{14}
which are depicted in \cite{20}. Hence we should expect cross sections from
that latter model similar to those obtained here with the current algebra
extrapolation.

In fig. 2 we show the two off shell extrapolations as a function of
\linebreak
$[-(p_{\pi}-q)^{2}]^{1/2}$  for $\vec{p}_{\pi} = 0, p^{0}_{\pi} = m_{\pi}$ and
$q^{0}=m_{\pi}/2$.
This simplified kinematics, appropriate at threshold where we study the
reaction, is used to evaluate the three dimensional integrals which appear
when we include the initial and final state interaction, as we shall discuss
bellow.

Chiral perturbation theory \cite{21} offers a natural framework to account
for, and extend, the results of current algebra, but still has problems in the
baryon sector \cite{22,23} and  is restricted to low momenta.
The $\pi N$ amplitude within this framework has been the subject of debate
\cite{24,25} but extensions of the method could prove most useful in
providing reliable off shell extrapolations in the region of interest to
this problem.

{}From ref. \cite{4,5,6} we know that taking into account initial and final
state interaction of the protons in this process near threshold is very
important.
In order to implement this we follow the standard procedure \cite{1} and
recall that the pion goes out in $s$-wave and the final state of the two
protons is $L=0, S=0, T=1, J=0$. Consequently the initial state is $J=0,
T=1, L=1, S=1$, because of angular momentum and parity conservation.
We work in coordinate space. Technically this requires to Fourier
transform the amplitude of eqs. (2), (3), which are given in momentum
space, and then perform integrals in coordinte space using the proper
initial and final $p p$ wave functions.
We use
radial wave functions solutions of the Schr\"odinger equation using the $NN$
interaction of the Paris \cite{26} and Bonn \cite{16} potentials.

The Coulomb interaction is known to decrease the cross section calculated
omitting it by an amount which ranges from
$30\%$ at $\eta = 0.2$ to $10\%$ at $\eta = 0.5
(\eta = p_{\pi max}/m_{\pi})$ \cite{4} \cite{27}. We implement these
corrections
to the results obtained using the strong potential alone.

In figs. 3 and 4 we show the result for the $pp \rightarrow pp \pi^{0}$ cross
sections close to threshold using the Paris and Bonn $NN$ potentials
respectively. As found in earlier work \cite{4,5,6} we see that using the on
shell $\pi N$ isoscalar amplitude leads to cross sections which are very
small compared with experiment. However, when the off shell extrapolation is
used, the cross section is considerably increased.
With the model of Hamilton we obtain
results close to experiment. With the current algebra
model for the $\pi N$ interaction
we obtain results which exceed the experimental cross section by about a
factor of two.

The differences between the results obtained with the Paris and Bonn $NN$
potentials in figs. 3 and 4 are mostly due to the differences appearing in the
Born term, which is about $50\%$ bigger with the Bonn $NN$ potential.

All this is telling us that the present knowledge of the off shell $\pi N$
extrapolation does not allow us to be very precise on the predictions for the
cross section of the $pp \rightarrow pp \pi^{0}$ reaction close to threshold,
but with present uncertainties it is clear that this off shell extrapolation
can by itself explain the experimental data.

It would be interesting to pursue research in the direction of ref. \cite{9}
to make the $NN, N\bar{N}, \bar{N}\bar{N}$ models more consistent with
experimental data on all these channels, or exploit theoretical models like
the one of ref. \cite{10}, with correlated two pion exchange, extrapolating
the model to negative energy states. This would bring new light into the
relativistic potentials which are the base of the previous theoretical
interpretation of the present problem \cite{6}. If the relativistic scalar
and vector potentials are smaller than  those used in \cite{6}, as claimed in
\cite{9}, or hinted with the use of correlated two pion exchange in the
negative energy sector, the off shell extrapolation of the $\pi N$ isoscalar
amplitude would stand as a likely candidate for the explanation of the
$pp \rightarrow pp \pi^{0}$ data. This should stimulate theoretical and
experimental work to put further constraints on the off shell extrapolation
of the $\pi N$ amplitude in order to reduce the uncertainties which we still
have in this cross section, as we have shown in this paper.

\vspace{2cm}

Acknowledgements:

We would like to thank useful discussions with S. Hirenzaki. This work has
been partly supported by CICYT, contract n$^{\circ}$. AEN 93-1205.
One of us E.H.
would
like to acknowledge finantial support from the Ministerio de Educaci\'on, in
the program of reincorporaci\'on de doctores.

\newpage

figure captions

fig. 1. Feynman diagrams considered in the $pp \rightarrow pp \pi^{0}$
reaction near threshold. a) Born term; b) rescattering term.

\vspace{1cm}

fig. 2. Off shell extrapolation of the $\pi N$ isoscalar amplitudes in the
Hamilton \cite{16} (solid line) and current algebra model (regularized at high
momenta)
\cite{18} as a function of $\sqrt{-t}$ (dashed-doted).

\vspace{1cm}

fig. 3. Cross section for $pp \rightarrow pp \pi^{0}$ near threshold as a
function of $\eta (p_{\pi max}/\mu )$ using the Paris $NN$ potential for the
initial and final state interaction of the two protons. Dashed line: Born
term and rescattering term with on shell $\pi N$ isoscalar amplitude. Solid:
with off shell $\pi N$ amplitude from Hamilton's model \cite{17}. Dashed doted
line: same with the current algebra $\pi N$ extrapolation \cite{18}. Coulomb
effects are included as discussed in the text.

\vspace{1cm}

fig. 4. Same as fig. 3 using the Bonn $NN$ potential for the initial and final
state interaction of the two protons


\newpage

\begin{thebibliography}{99}
\bibitem{1} D.S. Koltun and A. Reitan, Phys. Rev. 141 (1966)1413.
\bibitem{2} G. H\"ohler, F.Kaiser, R. Koch and E. Pietarinen, Handbook of
Pion-Nucleon Scattering [Physics Data 12-1 (1979)].
\bibitem{3} H. O. Meyer et al., Phys. Rev. Lett 65 (1990)2846, Nucl. Phys.
A539 (1992) 633.
\bibitem{4} G.A. Miller and P.U. Sauer, Phys. Rev. C44 (1991) R1725.
\bibitem{5} J.A. Niskanen, Phys. Lett. B289 (1992) 227.
\bibitem{6} T.S.H. Lee and D.O. Riska, Phys. Rev. Lett. 70(1993) 2237.
\bibitem{7} P.G. Blunden and D.O. Riska, Nucl. Phys. A536 (1992) 697.
\bibitem{8} K. Tsushima, D.O. Riska and P.G. Blunden, Nucl. Phys. A 559 (1993)
543. This paper corrects some of the relevant formulae of \cite{7}.
\bibitem{9} J.A. Tjon and S.J. Wallace, Phys. Rev. C32(1985) 1667; ibid C35
(1987)280; ibid C36 (1987) 1085.
\bibitem{10} H.C. Kim, J.W. Durso and K. Holinde, Phys. Rev. C49 (1994) 2355.
\bibitem{11} C. Garc\'{\i}a-Recio, E. Oset and L.L. Salcedo, Phys. Rev. C 37
(1988) 194.
\bibitem{12} D. Ashery and J.P. Schiffer, Ann. Rev. Nucl. Part. Sci.
36 (1986) 207; H. J. Weyer, Phys. Rep. 195 (1990) 295; G. Backenstoss, Phys.
Rep. 225 (1993) 97.
\bibitem{13} W.T. Huang, C.A. Levinson and M.K. Banerjee, Phys. Rev. C5
(1972)651.
\bibitem{14} C. Sch\"utz, J.W. Durso, K. Holinde and J. Speth, Phys. Rev. C49
(1994)2671.
\bibitem{15} F. Mandl and G. Shaw, Quantum Field Theory, John Wiley and Sons,
1984.
\bibitem{16} R. Machleidt, Advances in Nucl. Phys. 19 (1989) 189 (non
relativistic model, potential A of table A.3).
\bibitem{17} G. Hamilton, High Energy Physics, Ed. E.H.S. Burhop, Vol. 1,
Academic Press, N.Y. 1967, p. 194.
\bibitem{18} S.A. Coon et al., Nucl. Phys. A317(1979)242
\bibitem{19} M.K. Banerjee and J.B. Cammarata, Phys. Rev. D18(1978)4078, and
references therein.
\bibitem{20} L.L. Salcedo, K. Holinde, E. Oset and C. Sch\"ultz, University
of Valencia preprint.
\bibitem{21} J. Gasser and H. Leutwyler, Ann. Phys. 158 (1984) 142.
\bibitem{22} U. G. Meissner, Rep. Prog. Phys. 56 (1993) 903.
\bibitem{23} A. Pich, Rep. Prog. Phys., in print.
\bibitem{24} T.E.O. Ericson, Proc. Int. Workshop on pions in nuclei,
Edts., E. Oset, M.J. Vicente-Vacas and C. Garc\'{\i}a-Recio, World
Scientific, 1992, pag. 619.
\bibitem{25} M.E. Sainio, as in [24], pag. 652.
\bibitem{26} M. Lacombe et al., Phys. Rev. C 21 (1980) 861.
\bibitem{27} H.O. Meyer, Acta Physica Polonica, 24 (1993) 1735.
\end{thebibliography}
\end{document}